\documentclass[twocolumn,superscriptaddress,aps]{revtex4-1}
\usepackage{amsmath}
\usepackage{amssymb}
\usepackage{graphicx}
\graphicspath{{../Figures/}}
\usepackage{color}
\usepackage[colorlinks=true,linkcolor=blue,citecolor=blue]{hyperref}

\usepackage{siunitx}
\usepackage{amsfonts}
\usepackage{bm}
\usepackage{euscript}
\usepackage{textcomp}
\usepackage{gensymb}
\usepackage{float}
\usepackage{enumitem}
\usepackage{xcolor}
\usepackage{mhchem}

\newcommand{\blutx}[1]{\textcolor{blue}{}}

\newcommand{\redtx}[1]{\textcolor{red}{#1}}

\newcommand{\rfig}[1]{Fig.~\ref{#1}}
\newcommand{\rFig}[1]{Figure~\ref{#1}}

\newcommand{\mub}{\mu_\text{B}}

\newcommand{\jmma}{J_{1}^\text{MM}}
\newcommand{\jmmb}{J_{2}^\text{MM}}
\newcommand{\jmmc}{J_{3}^\text{MM}}
\newcommand{\jmmk}{J_{k}^\text{MM}}
\newcommand{\jmt}{J^\text{MT}}

\begin{document}

\title{Low temperature competing magnetic energy scales in the topological ferrimagnet TbMn$_{6}$Sn$_{6}$}

\author{S. X. M. Riberolles}
\affiliation{Ames Laboratory, Ames, IA, 50011, USA}

\author{Tyler J. Slade}
\affiliation{Ames Laboratory, Ames, IA, 50011, USA}
\affiliation{Department of Physics and Astronomy, Iowa State University, Ames, IA, 50011, USA}

\author {D.~L.~Abernathy}
\affiliation{Oak Ridge National Laboratory, Oak Ridge, TN 37831 USA}

\author {G.~E.~Granroth}
\affiliation{Oak Ridge National Laboratory, Oak Ridge, TN 37831 USA}

\author{Bing Li}
\affiliation{Ames Laboratory, Ames, IA, 50011, USA}
\affiliation{Department of Physics and Astronomy, Iowa State University, Ames, IA, 50011, USA}

\author{Y. Lee}
\affiliation{Ames Laboratory, Ames, IA, 50011, USA}

\author{P.~C.~Canfield}
\affiliation{Ames Laboratory, Ames, IA, 50011, USA}
\affiliation{Department of Physics and Astronomy, Iowa State University, Ames, IA, 50011, USA}

\author{B. G. Ueland}
\affiliation{Ames Laboratory, Ames, IA, 50011, USA}

\author{Liqin Ke}
\affiliation{Ames Laboratory, Ames, IA, 50011, USA}

\author{R.~J.~McQueeney}
\affiliation{Ames Laboratory, Ames, IA, 50011, USA}
\affiliation{Department of Physics and Astronomy, Iowa State University, Ames, IA, 50011, USA}

 \date{\today}

\begin{abstract}
TbMn$_{6}$Sn$_{6}$ is a metallic ferrimagnet displaying signatures of both topological electrons and topological magnons arising from ferromagnetism and spin-orbit coupling within its Mn kagome layers. Inelastic neutron scattering measurements find strong ferromagnetic (FM) interactions within the Mn kagome layer and reveal a magnetic bandwidth of $\sim230$ meV.  The low-energy magnetic excitations are characterized by strong FM Mn-Mn and antiferromagnetic (AFM) Mn-Tb interlayer magnetic couplings. We observe weaker, competing long-range FM and AFM Mn-Mn interlayer interactions similar to those driving helical magnetism in the YMn$_{6}$Sn$_{6}$ system. Combined with density-functional theory calculations, we find that competing Mn-Mn interlayer magnetic interactions occur in all $R$Mn$_6$Sn$_6$ compounds with $R=$ Y, Gd$-$Lu, resulting in magnetic instabilities and tunability when Mn-$R$ interactions are weak. In the case of TbMn$_{6}$Sn$_{6}$, strong AFM Mn-Tb coupling ensures a highly stable three-dimensional ferrimagnetic network.
\end{abstract}

\maketitle

\section{Introduction}
The potential technological applications of magnetic topological insulators and Weyl semimetals has generated new research directions aimed at understanding the coupling between magnetism and topological fermions.  This has brought renewed interest in magnetic kagome metals, such as Mn$_3$Ge \cite{Kiyohara16, Nayak16}, Fe$_3$Sn$_2$ \cite{Kida11, Lin18}, Co$_3$Sn$_2$S$_2$\cite{Lin12,Liu18, Yin19} and FeSn \cite{Kang20}, where both magnetism and topological electronic band crossings are hosted in the kagome layer.  Interesting topological responses, such as large anomalous Hall conductivity, are tied to the underlying magnetic order that can be impacted by both geometrical frustration and Dzyaloshinskii-Moriya (DM) interactions. In principle, these materials may host topological magnons in the presence of DM interactions \cite{Mook14}, opening up even more interesting avenues for the study of topological phenomena in metallic kagome systems. 

The hexagonal $R$Mn$_6$Sn$_6$ ($R$166) compounds ($R=$ rare-earth) consist of alternating Mn kagome and $R$ triangular layers. Contemporary studies of $R$166 compounds have focused on the interplay between their complex magnetism and topological electronic kagome band crossings \cite{Yin20,Ghimire20,Li21,Dhakal21,Wang21,Dally21,Mielke21,Zhang20}. $R$166 materials display a variety of magnetic structures, including antiferromagnetic (AFM), ferrimagnetic, and complex helical ordering, that are dependent on the nature of the host $R$ ion \cite{chafik91,Elidrissi91, Venturini91, Venturini96, Malaman_1997, Malaman99, Mazet_1999, Ghimire20,Lefevre_2002}. In addition, unique temperature and field-driven magnetic instabilities found in $R$166 compounds \cite{Clatterbuck99, Zaikov00, Dally21, Wang21} promise to open new avenues in topological state control and switching.

In $R$166, the intralayer Mn-Mn interactions are strongly ferromagnetic (FM) and magnetic complexity arises from a combination of competing Mn and $R$ magnetic anisotropies (for moment-bearing $R$-ions) and competing interlayer magnetic interactions \cite{Zajkov00, Irkhin02, Guo08, Rosenfeld08}.  $R$166 compounds with non-moment-bearing rare-earths, such as Y166, are easy-plane AFMs where competing FM and AFM coupling between FM Mn layers drives transitions from collinear to complex helical magnetic phases displaying net chirality and a topological Hall response in applied magnetic fields \cite{Venturini96, Ghimire20, Dally21, Wang21}. For moment-bearing rare-earths, the magnetism is strongly affected by rare-earth anisotropy and coupling between Mn and $R$ layers.  In Tb166, strong uniaxial anisotropy of the Tb ion and AFM Mn-Tb coupling favors unique uniaxial collinear ferrimagnetic state that has realized Chern-gapped topological fermions with a quantized magnetotransport response \cite{Yin20}. Surprisingly, Tb166 possesses a spin reorientation transition from easy-axis to easy-plane ferrimagnetism \cite{Malaman99,chafik91,Clatterbuck99, Zaikov00}. These discoveries demonstrate great potential for novel topological phenomena to be discovered by exploring other $R$166 materials via rare-earth engineering \cite{Ma21} or by the application of symmetry-breaking external fields.

To access this potential, we must address several open questions regarding the fundamental nature of the magnetism within $R$166 compounds. For example, is the Mn magnetism of an itinerant or local-moment nature and are the Mn-Mn interactions transferrable across the $R$166 materials? What is the variability of $R$-Mn interactions and $R$ anisotropy across the series?  Also, given recent reports on the connection between thermally driven magnetic fluctuations and quantum transport in Y166 \cite{Ghimire20} and Tb166 \cite{Mielke21}, what is the role of magnetic fluctuations in the emergent topological properties through the $R$166 family?

Here, we address the magnetic interactions in Tb166 in detail using inelastic neutron scattering (INS) and density-functional theory (DFT) calculations.  Using INS, we observe a hierarchy of competing interlayer Mn-Mn interactions in Tb166 similar to those used to explain the complex temperature- and field-driven helical magnetism observed in Y166 \cite{Venturini96, Ghimire20, Dally21,Zhang20, Wang21}. We find that strong uniaxial Tb magnetic anisotropy and AFM coupling between Mn and Tb layers generates a rigid three-dimensional ferrimagnetic lattice. A clean spin gap of 6.5 meV suppresses collective spin fluctuations at temperatures relevant for quantum transport ($<$ 20 K). Thus, it is likely that the main avenue available for tuning the topological band states in Tb166 is by controlling the spin reorientation transition. Results of our DFT calculations largely agree with the sign, magnitude and overall hierarchy of interlayer couplings found experimentally after the introduction of on-site Coulomb repulsion (DFT+U).

The INS data also show that FM intralayer Mn-Mn interactions in both Tb166 and Y166 (Ref.~\cite{Zhang20}) are comparably strong and push the overall magnon bandwidth up to $\sim$230 meV.  However, increasingly broad lineshapes for Tb166 do not allow the observation of magnetic excitations above $\sim$ 125 meV. Unlike reports of a K-point gap caused by DM interactions in the magnon spectrum of Y166 \cite{Zhang20}, this severe line broadening in Tb166 obscures any evidence of a topological magnon gap. This suggests that, despite our quantitative modeling of the spin-wave spectrum presented here, there is still much to be learned about the itinerant character of Mn magnetism and the role of spin-orbit interactions in $R$166 materials.

 \section{Experimental Details}
\begin{figure*}
\includegraphics[width=1.0\linewidth]{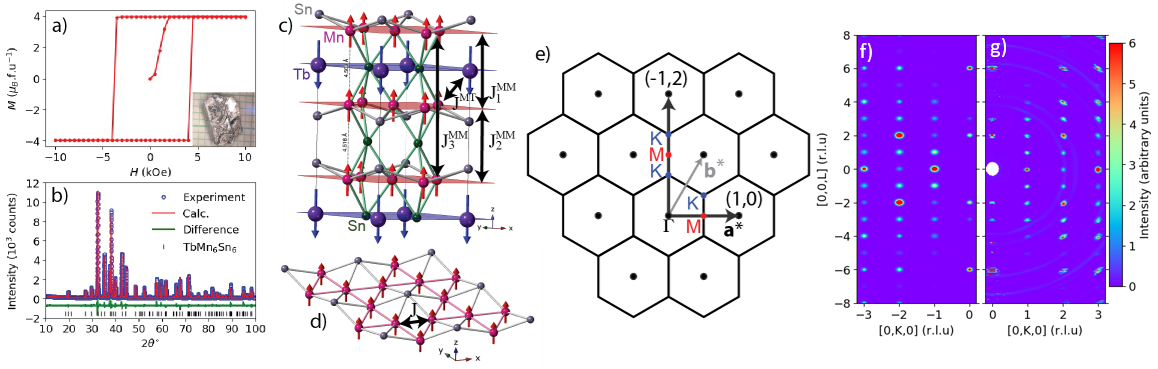}
\caption{\footnotesize \redtx{} 
(a) Single-crystal magnetization data for Tb166 recorded at 2 K with \textit{H} applied along \textit{c}.  The inset shows a typical  single-crystal sample of Tb166. (b) Powder x-ray diffraction measurements of Tb166 collected at room temperature and fitted using Rietveld refinement analysis. 
(c) Ferrimagnetic ground state structure of TbMn$_6$Sn$_6$.  Key interlayer interactions are shown with heavy black arrows.  
(d) Magnetic interactions within a single Mn-Sn kagome layer. 
(e) 2D hexagonal Brillouin zone showing conventional reciprocal lattice vectors $\textbf{a}^{*}$ and $\textbf{b}^{*}$ and special points, $\Gamma$ (black), K (blue) and M (red). Inelastic neutron scattering data are discussed in terms of the orthogonal vectors (1,0) and (-1,2).
(f) Simulated (0,$K$,$L$) elastic single crystal neutron scattering intensity containing both nuclear and magnetic components for Tb166 below 250 K. The reciprocal space is here set in the conventional way. Antiparallel magnetic moments of 9.0(Tb) and 2.17(Mn) $\mu_{\text{B}}$ are set along $c$.  
(g) Tb166 elastic single crystal neutron scattering data collected on ARCS in the (0,$K$,$L$) scattering plane at 7 K.}
\label{Char_plus_struct}
\end{figure*}

Single crystals of Tb166 were grown from excess Sn using the flux method. A nominal (TbMn$_6$)$_5$Sn$_{95}$ molar ratio of elemental Tb (Ames Laboratory 99.9\%), Mn (Research Organic/Inorganic Chemical Corp, 99.995\%), and Sn (Alfa Aesar, 99.99\%) was weighed and loaded into the growth side of a 5 mL fritted alumina crucible set~\cite{Canfield_2016}. The crucibles were flame sealed under vacuum inside an 18 mm diameter fused silica ampule with a small amount of silica wool placed above and below the crucibles to serve as cushioning, and heated to 1180$^\circ$C in 12 hours.  After dwelling at 1180$^\circ$C for 3 hours, the furnace was quickly cooled in 3 hours to 775$^\circ$C and then slowly cooled over 300 hours to 575$^\circ$C. Upon reaching the final temperature, the tube was rapidly removed from the furnace, inverted into a metal centrifuge, and the excess flux decanted. The crucibles were opened to reveal large (up to 300 mg), shiny, hexagonal crystal plates (see Fig.~\ref{Char_plus_struct}(a)).

Low temperature magnetization was measured using a Quantum Design Magnetic Property Measurement System (MPMS 3), SQUID magnetometer  ($T= 1.8-300$ K, $H_{max} = $70 kOe). A Tb166 single crystal sample was mounted on a plastic disc and the field was applied along \textit{c}. Prior to measuring the sample, the blank disc was measured and used for a background subtraction. Figure~\ref{Char_plus_struct}(a) shows low temperature magnetization measured at 2~K with $H \parallel c$ that accurately reproduce the previously reported hysteresis loop displaying a saturated magnetization of $\approx$~4~$\mu_{\text{B}}$/f.u.~\cite{Kimura_2005}.
    
Tb166 crystallizes in the HfFe$_6$Ge$_6$-type structure with hexagonal space group \textit{P6/mmm} (No.~191) and Mn, Sn1, Sn2, Sn3 and Tb ions, respectively, sitting at the 6i, 2e, 2d, 2c and 1b Wyckoff positions~\cite{Olenitch_1991}, see Fig.~\ref{Char_plus_struct}(c-d). From a Rietveld analysis of XRD data collected at 300~K (see Fig.~\ref{Char_plus_struct}(b)), we obtain refined values of 5.53317(6) and 9.0233(1)~\AA~for lattice parameters $a$ and $c$, as well as atomic coordinates z$_{\mathrm{Mn}}$=0.2539(2) and z$_{\mathrm{Sn1}}$=0.1624(2), in close agreement with previous reports~\cite{Malaman99,chafik91}. Below 423~K, both the Mn and Tb layers simultaneously develop FM order, but couple antiferromagnetically, resulting in an overall ferrimagnetic order. All magnetic moments initially lie in the basal plane, but remarkably, upon cooling between 350~K and 305~K a spin reorientation takes place, resulting in the ground state collinear ferrimagnetic arrangement of Mn and Tb moments along the $c$-axis~\cite{chafik91, Malaman99} shown in Fig.~\ref{Char_plus_struct}(c) and \ref{Char_plus_struct}(d). 

INS measurements were performed on the Wide Angular-Range Chopper Spectrometer (ARCS) located at the Spallation Neutron Source at Oak Ridge National Laboratory~\cite{Abernathy_2012}. An array of five crystals with a total mass of 495.6 mg was co-aligned with the ($H,0,L$) scattering plane set horizontally, and attached to the cold head of a closed-cycle-refrigerator. Data were collected at the base temperature of 7 K using incident energies of $E_i =$ 30, 75, 160 and 250 meV [elastic resolutions are listed in Table I in the Supplemental Material (SM) \cite{SI}]. For $E_i =$ 30, 160 and 250 meV, the sample was rotated around 180 degrees in one degree increments for full coverage of \textbf{q}, $E$ space, where \textbf{q} ($E$) is the momentum (energy) transfer, respectively. For $E_i =$ 75 meV, the rotation increment was reduced to half a degree.

The INS data were reduced to \textbf{q} and $E$, symmetrized to improve statistics, and cuts made for further analysis using Mantid~\cite{Mantid}. The neutron scattering data are described using the momentum transfer in hexagonal reciprocal lattice units, ${\bf q}(H,K,L) = \frac{2\pi}{a}\frac{2}{\sqrt{3}}(H\hat{a}^*+K\hat{b}^*)+\frac{2\pi}{c}L\hat{c}$. The INS data are presented in terms of the orthogonal vectors $(1,0,0)$, $(-1,2,0)$, and (0,0,1), as shown in Fig.~\ref{Char_plus_struct}(e). Special K- and M-points in the Brillouin zone are found at $(H,K,L) = (\frac{1}{3},\frac{1}{3},0)$ and ($\frac{1}{2}$,0,0) and symmetry-related points, respectively. The INS data are displayed as intensities that are proportional to the spin-spin correlation function $S({\bf q},E)$. To improve statistics, the data have been symmetrized with respect to the crystallographic space group \textit{P6/mmm}.

We first examined the elastic scattering from our co-aligned crystals [shown in Fig.~\ref{Char_plus_struct}(g)] and compared the data to simulations of the nuclear and magnetic scattering [shown in Fig.~\ref{Char_plus_struct}(f)].  Using the Bilbao crystallographic server, we find that below 250 K the magnetic structure adopts the high-symmetry magnetic space group \textit{P6/mm'm'} (No.~191.240) where both magnetic sublattices are restricted to have their ordered moments lying along the \textit{c}-axis \cite{bilbao}. The ordered magnetic moment at 4.5~K are reported as 2.17 and 9.0 $\mu_{\text{B}}$ for Mn and Tb, respectively \cite{Clatterbuck99}. Using these values and the \textit{P6/mm'm'} symmetry, we simulated the corresponding nuclear and magnetic neutron diffraction patterns for the ($0,K,L$) plane using \textsc{mag2pol}  \cite{mag2pol}. The good agreement obtained between simulated and experimentally measured patterns confirms the high quality of our samples as well as the previously reported low temperature ferrimagnetic ground state in Tb166.

\section{Minimal Heisenberg model for the spin excitations}
\begin{figure}
\includegraphics[width=1.0\linewidth]{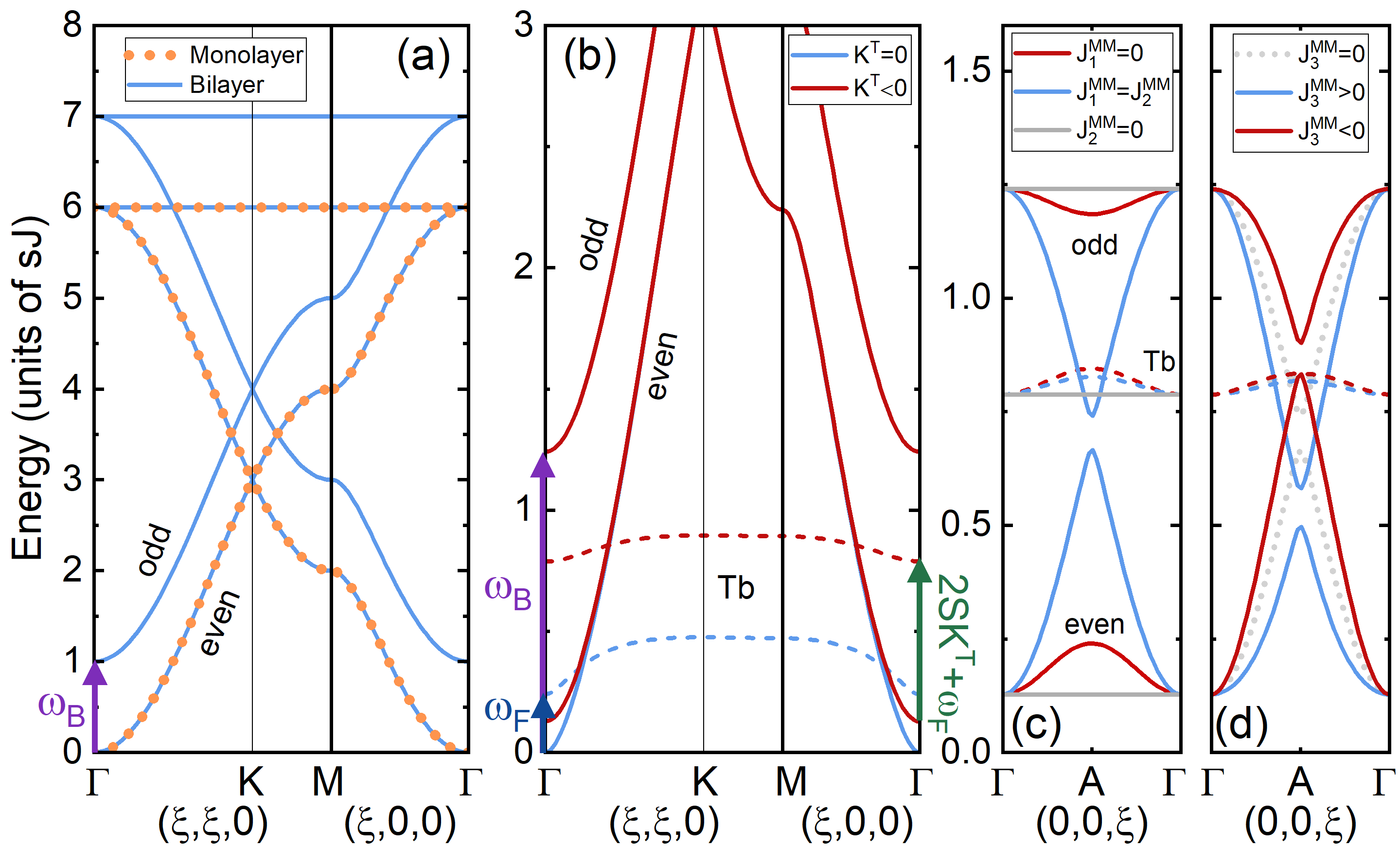}
\caption{\footnotesize (a) Monolayer kagome spin wave dispersion with energy in units of $s|J|$ (orange dots) and Mn-Mn bilayer dispersion with $J^{\rm MM}_2=0.5|J|$ (blue lines).  The latter shows the bilayer splitting of odd and even modes by $\omega_B=2s|J^{\rm MM}_2|$.  (b) Low-energy dispersion when Mn bilayers are coupled through Tb with $S=3s$ and  $J^{\rm MT}=-0.04J$ (blue lines).  The odd bilayer mode and Tb mode (dashed line) are shifted by the ferrimagnetic exchange field, $\omega_F=2(6s-S)J^{\rm MT}$, as shown.  Red lines include uniaxial Tb single-ion anisotropy with $K^{\rm T}=0.07J$ and $K^{\rm M}=0$ that introduces a spin gap in the even mode ($\Delta$) and increases the Tb  mode spin gap ($\Delta_{\mathrm{Tb}}$). (c) Interlayer dispersion of low-energy branches with identical bilayer splitting, $J^{\rm MM}_1+J^{\rm MM}_2=0.5J$ for cases where $J^{\rm MM}_1=J^{\rm MM}_2$ (blue lines), $J^{\rm MM}_1=0$ (red lines), and $J^{\rm MM}_2=0$ (gray lines).  (d)  Interlayer dispersion of low-energy branches when $J^{\rm MM}_1=J^{\rm MM}_2$ and the coupling between Mn layers in adjacent unit cells, $J^{\rm MM}_3$, is either ferromagnetic (red lines), antiferromagnetic (blue lines) or zero (gray dashed lines).}
\label{fig_disp_schematic}
\end{figure} 

Before describing the INS data, we first discuss a minimal description of the magnetic interactions in Tb166 and the key features of the resultant spin excitations.  Kagome layers are known for unusual magnetic behavior due to geometric frustration and the role of spin-orbit coupling via the DM interaction. 
All known hexagonal $R$166 compounds possess FM kagome layers with an easy-plane Mn magnetic anisotropy which minimizes the role of intralayer geometric frustration \cite{Baranov11}. However, the competition between Mn-Mn FM and AFM interlayer magnetic interactions is known to cause magnetic instabilities in Y166 that lead to complex helical phases \cite{Venturini96,Zhang20, Ghimire20}.  

For $R$166 compounds with magnetic rare-earth ions, two additional factors control the magnetic behavior. The first is strong AFM coupling between the $R$ and Mn sublattices that can result in tightly bound Mn-$R$-Mn collinear ferrimagnetic trilayers. The second factor is the single-ion anisotropy of the rare-earth ion.  For ferrimagnetic Gd166, the weak anisotropy of the spin-only Gd$^{3+}$ ion combined with easy-plane Mn anisotropy and Gd$-$Mn AFM coupling results in antiparallel ordered Gd and Mn moments lying in the basal layer~\cite{Malaman99}.  On the other hand, $R=$ Tb$-$Ho ions possess uniaxial anisotropy that competes with the Mn easy-plane anisotropy.  This competition, along with higher-order contributions to the $R$ anisotropy\cite{Zajkov00, Guo08}, drives spin reorientation transitions where the ordered Mn and $R$ moments rotate in unison~\cite{Malaman99,chafik91}. As mentioned above, Tb166 adopts an out-of-plane uniaxial ferrimagnetic ground state [see Fig.~\ref{Char_plus_struct}(c)], with Mn and Tb moments collectively rotating to fully lie in the basal plane above $T_{\mathrm{sr}}=$ 350 K.   $R=$~Dy and Ho are similar ferrimagnets with spin reorientation transitions, but the weaker $R$-ion anisotropy results in a ground state easy-axis that is tilted away from the $c$-axis \cite{Malaman99,chafik91}.  Close to $T_{\mathrm{sr}}$, the competing $R$ and Mn single-ion anisotropies drive first-order magnetization processes in applied magnetic fields \cite{Clatterbuck99, Zajkov00}.

Given the already interesting role of competing interlayer interactions in Y166 and competing anisotropies in $R=$ Tb$-$Ho, it remains to consider their combined role in $R$166 with magnetic rare-earths.  We define a general Heisenberg model with the Hamiltonian $\mathcal{H=H_{\rm intra}+H_{\rm inter}+H_{\rm aniso}+H_{\rm DM}}$ that consists of isotropic intralayer and interlayer pairwise exchange, single-ion anisotropy, and DM interactions.

In our minimal description, each Mn kagome layer possesses strong nearest-neighbor (NN) FM exchange ($J<0$) which determines the large overall magnon bandwidth. 
\begin{equation}
\mathcal{H}_{\rm intra}=J\sum_{\langle i<j \rangle} \bold{s}_i \cdot \bold{s}_j
\label{Hintra}
\end{equation}
Here, $\bold{s}$ is the Mn spin operator with magnitude $s=1$.  Fig.~\ref{fig_disp_schematic}(a) shows the dispersion for a single Mn kagome layer given by $\mathcal{H}_{\rm intra}$ within linear spin wave theory.  The overall bandwidth is $6s|J|$ with a Dirac band crossing at the K-point with energy $3s|J|$.  As described below, our data analysis does not benefit from the introduction of longer-ranged intralayer interactions, although we cannot exclude them. 

To describe the interlayer interactions, Fig.~\ref{Char_plus_struct}(c) shows that nearly equidistant FM Mn kagome layers are stacked along the $c$-axis. Tb layers are inserted after every two Mn layers and with opposite magnetization, forming a Mn($\uparrow$)-Mn($\uparrow$)-Tb($\downarrow$)-Mn($\uparrow$)-Mn($\uparrow$)-Tb($\downarrow$) pattern. Several unique interlayer magnetic couplings between Mn layers and between Mn and Tb layers are possible, giving
\begin{equation}
\mathcal{H}_{\rm inter}=\sum_{k} \sum_{i<j} J^{\rm MM}_k\bold{s}_i \cdot \bold{s}_{j+k} + J^{\rm MT} \sum_{\langle i<j \rangle} \bold{s}_i \cdot \bold{S}_j.
\label{Hinter}
\end{equation}
Here, $J^{\rm MT}>0$ is the AFM coupling between neighboring Mn and Tb layers, with Tb having a spin angular momentum of $S=3$. We label interactions between Mn layers by a layer index $k$ ($J^{\rm MM}_k$).  Due to the Tb layer, adjacent Mn layers above and below a given Mn layer are inequivalent.  Our data indicate that the FM coupling between next-nearest neighbor (NNN) Mn-Mn layers separated by a Sn$_4$ block ($J^{\rm MM}_2$) is stronger than the coupling between NN Mn-Mn layers separated by a TbSn$_2$ block ($J^{\rm MM}_1$), in agreement with analysis of neutron diffraction data \cite{Venturini96, Rosenfeld08}. 

By itself, $J^{\rm MM}_2$ forms strongly coupled FM Mn-Mn bilayers and generates a bilayer splitting $\omega_B=2s|J^{\rm MM}_2|$ of the single-layer dispersion into odd and even modes, as shown in Fig.~\ref{fig_disp_schematic}(a). The K-point splits into two (odd and even) topological magnon crossings that remain ungapped in the absence of DM interactions.  

The strong AFM interaction $J^{\rm MT}$ generates a ferrimagnetic exchange field with energy scale $\omega_F=2(6s-S)J^{\rm MT}$. $\omega_F$ increases the odd-even splitting and gives rise to a new branch of Tb character with a spin gap of $\Delta_{\rm Tb}=\omega_F$ at the $\Gamma$-point, as shown in Fig.~\ref{fig_disp_schematic}(b). 
 
The introduction of uniaxial single-ion anisotropy for both Tb and Mn ($K^{\rm T}$ and $K^{\rm M}$) is given by
\begin{equation}
\mathcal{H}_{\rm aniso}= K^{\rm M} \sum_{i} (s_i^z)^2 + K^{\rm T} \sum_{i} (S_i^z)^2
\label{Haniso}
\end{equation}
where the sums are over each sublattice.  Whereas Mn is expected to have a weak easy-plane anisotropy ($K^{\rm M} \gtrsim 0$), Tb has a large uniaxial anisotropy at low temperatures ($K^{\rm T}<0$).  With $K^{\rm M}=0$, $K^{\rm T}$ generates a spin gap $\Delta\approx\sqrt{2sSK^{\rm T}J^{\rm MT}}$ for the even branch and increases $\Delta_{\rm Tb}$ such that $\Delta_{\rm Tb}-\Delta=2SK^{\rm T}+\omega_F$, as shown in Fig.~\ref{fig_disp_schematic}(b).

We now consider the effect of $J^{\rm MM}_1$.  When $J^{\rm MM}_1=0$, the interlayer dispersion of the low-energy branches is mainly controlled by $J^{\rm MT}$.  As $J^{\rm MM}_1$ is increased, models indicate that the bilayer splitting becomes $\omega_B=2s|J^{\rm MM}_1+J^{\rm MM}_2|$ and the interlayer bandwidth of odd and even modes sharply increases and reaches a maximum when $J^{\rm MM}_1=J^{\rm MM}_2$, as shown in Fig.~\ref{fig_disp_schematic}(c). The limit where $J^{\rm MM}_2=0$ corresponds to isolated trilayer Mn-Tb-Mn blocks where the interlayer bandwidth is zero.  

To better describe the experimental data, an interaction between like Mn layers in adjacent unit cells, $J^{\rm MM}_3$, is introduced as well. As shown in Fig.~\ref{fig_disp_schematic}(d), $J^{\rm MM}_3$ oppositely affects the interlayer odd and even bandwidths while preserving the A-point gap at $\bold{q}= (0,0,1/2)$.  For example, when $J^{\rm MM}_3$ is AFM, the bandwidth of the odd mode increases and the even mode decreases. 

Finally, the presence of DM interactions is principally associated with gapping at the Dirac points at K and has recently been reported in Y166 \cite{Zhang20}. However, as described below, we find no clear evidence for a K-point gap in Tb166, due to the presence of strong damping.  Therefore, it is not necessary to introduce DM interactions to model our data ($\mathcal{H}_{\rm DM}=0$). 

\section{Interlayer dispersions}
\begin{figure}
\includegraphics[width=1.0\linewidth]{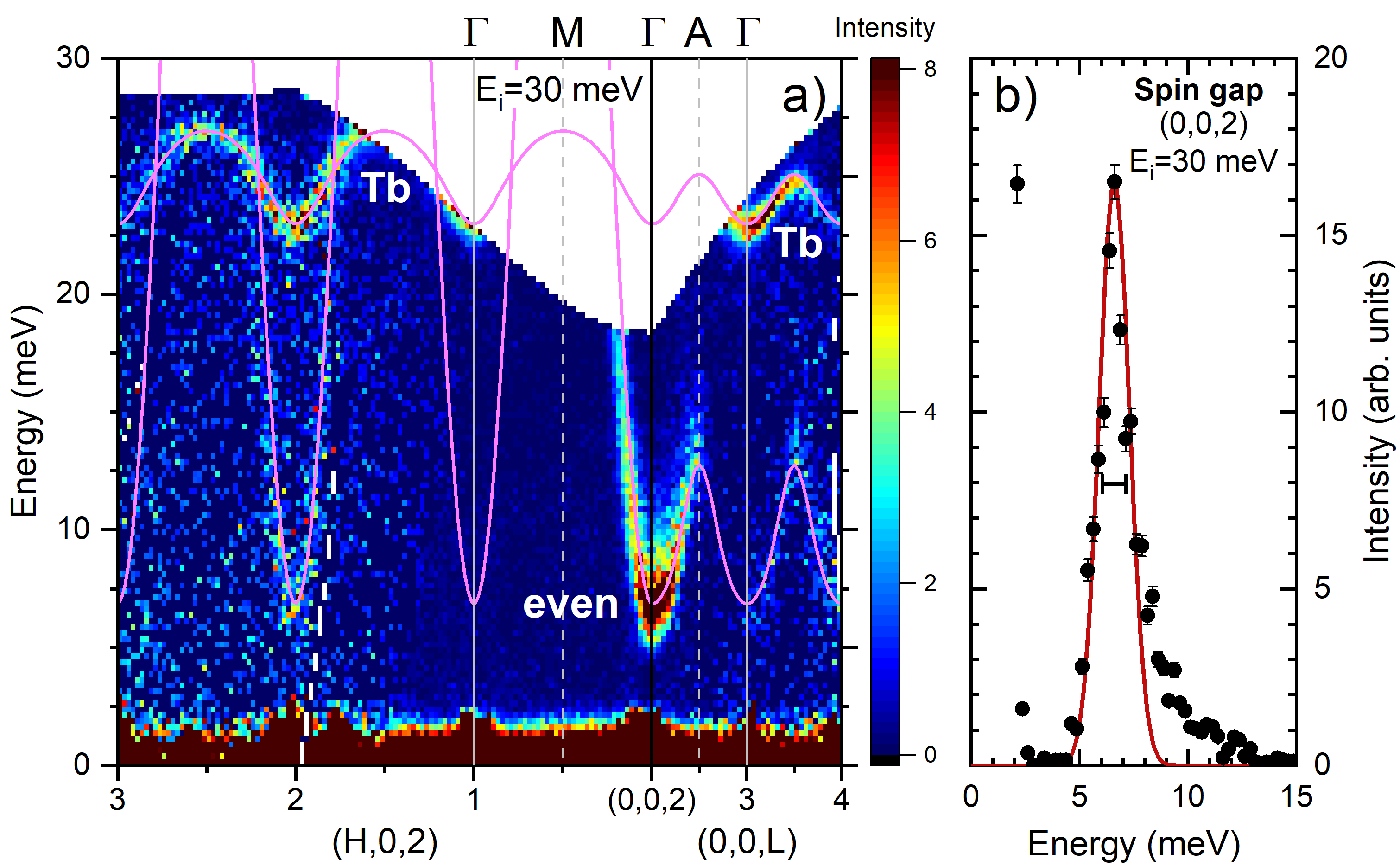}
\caption{\footnotesize (a) Slices of the neutron intensity showing the dispersion through the (0,0,2) $\Gamma$-point along $(H,0,0)$ and $(0,0,L)$ for data taken with $E_i=30$ meV.  Pink lines correspond to the model dispersion relation obtained from fits described below. Gray vertical lines identify Brillouin zone centers (solid) and zone boundary points (dashed), as labeled on the top axis.   (b) Energy spectrum through (0,0,2) averaged over $\textbf{q}$ ranges of $\Delta H=\Delta K= \pm 0.035$ and $\Delta L=0.1$ rlu.  The red line is a Gaussian fit that indicates a resolution-limited peak corresponding to a spin gap of $\Delta = 6.5$ meV.}
\label{fig_spin_gap}
\end{figure}

Having outlined the various expectations for the spin wave dispersion in Tb166, we now describe the features of the INS data. Figure \ref{fig_spin_gap}(a) shows a slice through the $E_i=$ 30 meV data along the $(H,0,0)$ and $(0,0,L)$ directions through the (0,0,2) $\Gamma$-point. The lowest energy mode is the even branch, which displays a clean spin gap of $\Delta=$ 6.5 meV as shown by the resolution-limited peak in the energy cut through the $\Gamma$-point at (0,0,2) [Fig.~\ref{fig_spin_gap}(b)].  Along $(0,0,L)$, the even branch has limited interlayer dispersion, reaching only 14 meV at the A-point, whereas the intralayer dispersion of the even branch along ($H$,0,0) extends to much higher energies. 

\begin{figure}
\includegraphics[width=1.0\linewidth]{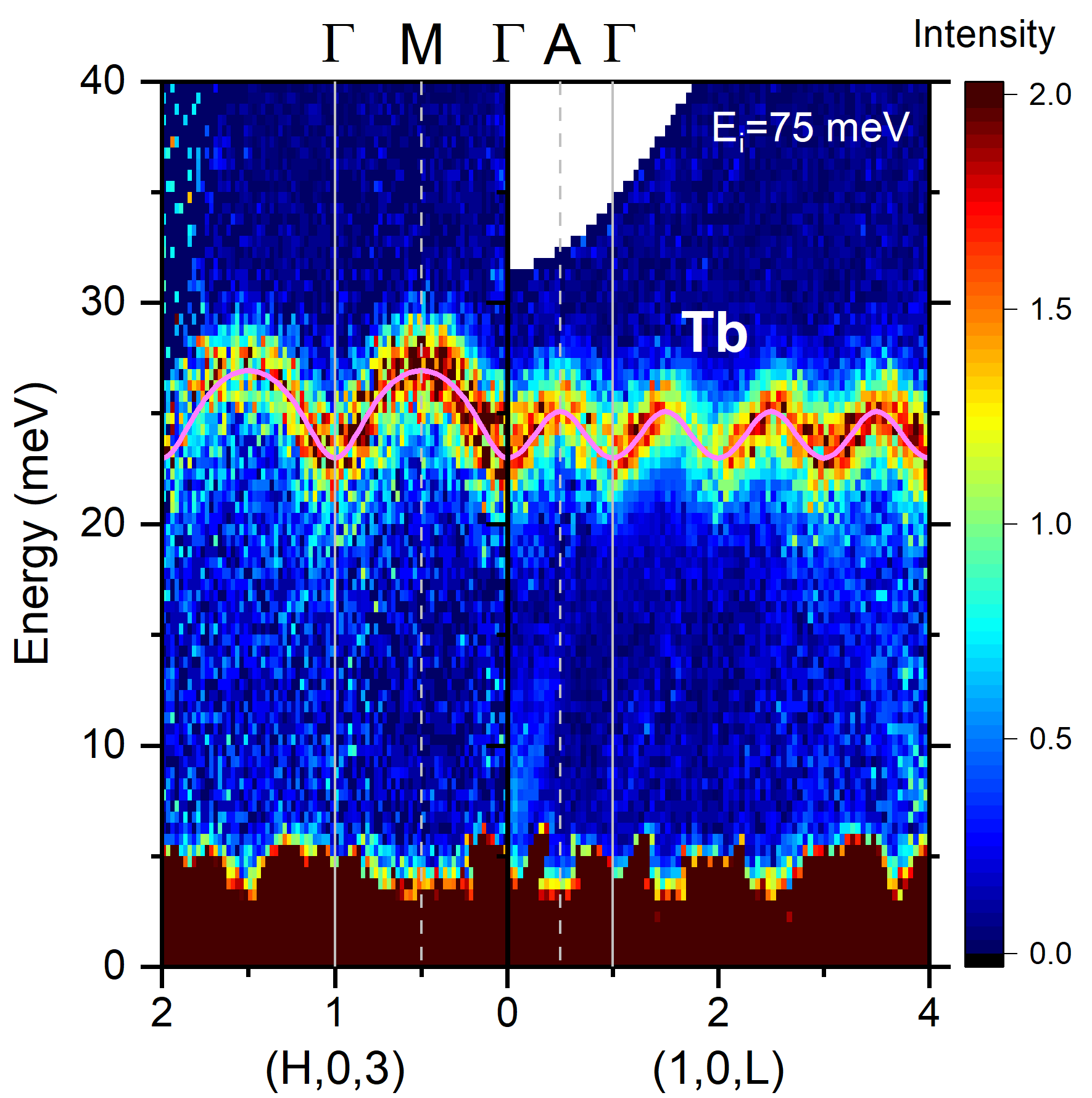}
\caption{\footnotesize Slices of the intensity along $(H,0,3)$ (left) and $(1,0,L)$ (right) with $E_i=$ 75 meV showing the intralayer and interlayer dispersion of the Tb mode, respectively. The two slices employed reciprocal space averaging of $\Delta L = \pm 0.1$ and $\Delta H= \pm 0.1$ rlu, respectively, with $\Delta K=\pm 0.058$ rlu used in both slices.  Pink lines correspond to the dispersion relation obtained from fits described below. Gray vertical lines identify Brillouin zone centers (solid) and zone boundary points (dashed), as labeled on the top axis.}
\label{fig_Tb_mode}
\end{figure}

We also glimpse a narrow band of excitations near $\sim$25 meV in Fig.~\ref{fig_spin_gap}(a) that corresponds to the Tb mode.  Figure \ref{fig_Tb_mode} shows the Tb mode dispersion along $(H,0,0)$ and $(0,0,L)$ more clearly using $E_i=$ 75 meV and focusing on Brillouin zones where the structure factor of the even branch is close to zero ($L=odd$ or $H=odd$).

\begin{figure}
\includegraphics[width=1.0\linewidth]{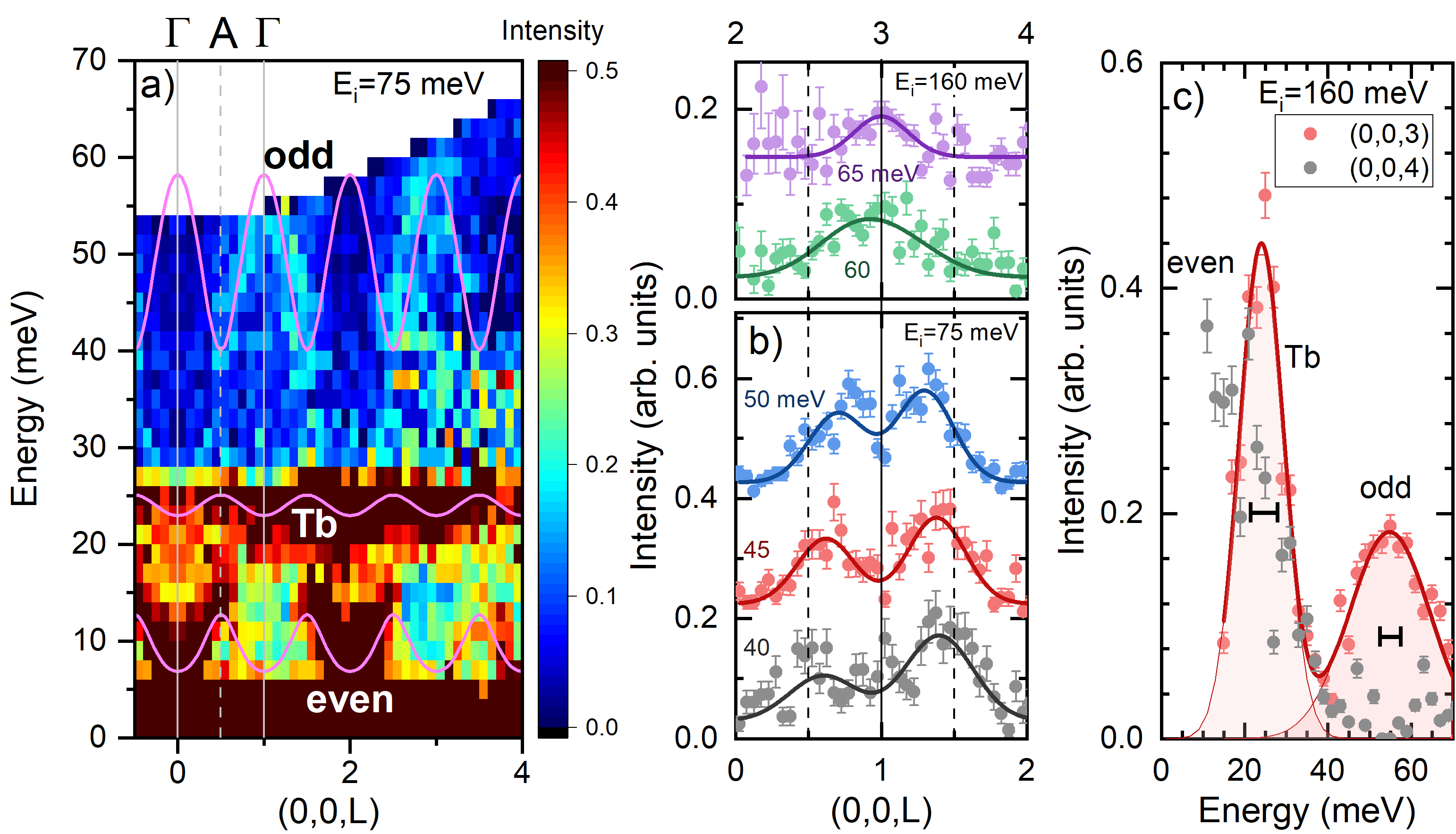}
\caption{\footnotesize (a) Slices of the intensity dispersion along $(0,0,L)$ with $E_i=$ 75 meV show the odd branch between $40-60$ meV. Pink lines correspond to the model dispersion relation obtained from fits described below. Gray vertical lines identify Brillouin zone centers (solid) and zone boundary points (dashed), as labeled on the top axis. (b) Constant energy cuts along $(0,0,L)$ at $E_i=75$ meV (lower panel) and 160 meV (upper panel) summed over $\Delta H= \pm 0.1$, $\Delta K= \pm0.058$ rlu and $\Delta E= \pm 2.5$ meV.  Gaussian fits reveal the dispersion of the odd branch. (c) Constant-${\bf q}$ cuts at (0,0,3) and (0,0,4) summed over $\Delta H= \pm 0.1$, $\Delta K= \pm0.058$, and $\Delta L=0.25$ rlu showing even, Tb, and odd modes at the $\Gamma$-point. }
\label{fig_odd_branch}
\end{figure}

The odd branch is observed in slices of the data taken with higher incident energies of 75 and 160 meV, as shown in Fig.~\ref{fig_odd_branch}.  The even and odd branches have structure factors that are maximized in Brillouin zones with $L= even$ and $L= odd$, respectively.  Fig.~\ref{fig_odd_branch}(a) and the constant energy cuts in Fig.~\ref{fig_odd_branch}(b) show that the interlayer odd branch disperses from roughly 60 meV at the $\Gamma$-point down to 40 meV at the A-point. Constant-$\bold{q}$ energy cuts at (0,0,3) and (0,0,4) in Fig.~\ref{fig_odd_branch}(c) also demonstrate a $\Gamma$-point energy of $\sim$60 meV for the odd branch. Considering the spin gap, this allows for an estimate of an odd-even splitting of $\omega_B+\omega_F \approx$ 55 meV.   Figs.~\ref{fig_odd_branch}(a)$-$(c) show that the odd branch is significantly weaker and broader than the resolution-limited low-energy even and Tb branches, but has a much larger interlayer bandwidth.

Various data cuts similar to those shown in Figs.~\ref{fig_spin_gap}$-$\ref{fig_odd_branch} were used to produce a list of dispersion points, $\omega_i(\bold{q})$, for even, odd, and Tb interlayer branches in various Brillouin zones. In this list, we also include the energies of the intralayer Tb modes along $(H,0)$ [Fig.~\ref{fig_Tb_mode}(a)] and $(-K,2K)$ whose dispersions are sensitive to $J^{\rm MT}$ and $K^{\rm T}$.  We used this list of 100 observables to fit the experimental dispersion to the reduced Heisenberg model $\mathcal{H=H_{\rm inter}+H_{\rm aniso}}$ using \textsc{SpinW} \cite{Toth15}.  The Mn and Tb spin values are fixed to $s=1$ and $S=3$, respectively.  

For $\mathcal{H}_{\rm aniso}$, the spin reorientation transition of Tb166 and the general magnetic structures of other $R$166 compounds suggest that Mn has weak easy-plane anisotropy ($K^{\rm M} \gtrsim 0$).  However, fixing $K^{\rm M}=0$ results in a fitted spin gap that is much lower than experimental values. We assume that this discrepancy is caused by additional contributions to the magnetic anisotropy, such as exchange anisotropy, that are not included in our model. The introduction of $K^{\rm M}<0$ to our fitting (as an effective uniaxial Mn anisotropy) dramatically improves the fitted spin gap. We note that alternative fitting schemes with $K^{\rm M}=0$ and anisotropic $J^{\rm MT}$ interactions give similar fitting results when $J^{\rm MT}_{zz} \approx 1.30J^{\rm MT}_{xx}$.  

For $\mathcal{H}_{\rm inter}$, the observed odd-even splitting of $\sim$55 meV is determined primarily by $|J^{\rm MM}_1+J^{\rm MM}_2|$ and the A-point gap of $\sim$25 meV by $|J^{\rm MM}_1-J^{\rm MM}_2|$.  However, the determination of the signs and relative strength of $J^{\rm MM}_1$ and $J^{\rm MM}_2$ requires careful fitting of the interlayer dispersions.  We ran 41 different fitting iterations starting with equal values of $J^{\rm MM}_1$ and $J^{\rm MM}_2$.  All fitting sessions find $J^{\rm MM}_1+J^{\rm MM}_2 \approx -24$ meV with two local minima where $J^{\rm MM}_2/J^{\rm MM}_1 \approx 4$ or 1/3.  Both interactions are FM.  The case where $J^{\rm MM}_2/J^{\rm MM}_1 \approx 4$ turns out to be the global minimum with a reduced $\chi^2=0.8$ which is lower than $\chi^2=1.0$ for the other case. The fits find that $J^{\rm MM}_2$ is the dominant interlayer interaction, confirming the expectation based on neutron diffraction studies of the double-flat spiral AFM structure of Y166 \cite{Venturini96,Zhang20, Ghimire20}. 

In the overall fits to $\mathcal{H}_{\rm inter}$, we find that an AFM $J^{\rm MM}_3$ must be introduced to account for the different bandwidths of even ($\sim$10 meV) and odd ($\sim$20 meV) interlayer dispersions, as shown in Figs.~\ref{fig_disp_schematic}(d) and \ref{fig_odd_branch}(a). An AFM $J^{\rm MM}_3$ will compete with FM $J^{\rm MM}_1$ and could lead to a destabilization of the ferrimagnetic stacking sequence.  However, calculations of the classical stability of the ferrimagnetic state described below suggest that  $J^{\rm MM}_3$ is not strong enough to create such an instability in Tb166.  Similar competing interactions have been proposed for Y166, but with AFM $J^{\rm MM}_1$ and FM $J^{\rm MM}_3$ \cite{Ghimire20,Dally21}.  This cannot be the case for Tb166, since the odd branch would have a minimum in the dispersion at $\Gamma$, which is not observed experimentally.
 
Fitting the spin wave dispersions produced the set of interlayer exchange parameters in Table \ref{Heisenberg} where error bars correspond to the variances obtained over all fitting iterations. Further details of the fitting procedure are described in the SM~\cite{SI}.  Within our model, the fit parameters predict an additional four modes (two odd and two even) at higher energies.  These modes are not clearly observed in the current experiment, as discussed below.

\begin{table}
\caption {Heisenberg parameters for TbMn$_6$Sn$_6$ as obtained from fits to the neutron data. }
\renewcommand\arraystretch{1.25}
\centering
\begin{tabular}{ c | c | c }
\hline\hline
~ Coupling ~ & ~Energy (meV)~ & ~ description ~  \\
\hline
$J$			& -28.8 (2)	& intralayer FM    \\
$J^{\rm MT}$		& 1.42 (6) 		& interlayer AFM    \\
$J^{\rm MM}_1$	& -4.4 (4)	& interlayer FM	    \\
$J^{\rm MM}_2$	& -19.2 (2) 	& interlayer FM   \\
$J^{\rm MM}_3$	& 1.8 (2) 	& interlayer AFM  \\
$K^{\rm M}$		& -1.30 (6) 	& uniaxial anisotropy  \\
$K^{\rm T}$		& -1.70 (12) 	& uniaxial anisotropy \\
$\omega_B$	& $\sim$47	& bilayer splitting \\
$\omega_F$	&  $\sim$ 8	& ferrimagnetic exchange \\
\hline\hline
\end{tabular}
\label{Heisenberg}
\end{table}

\section{Intralayer dispersions}

\begin{figure}
\includegraphics[width=1.0\linewidth]{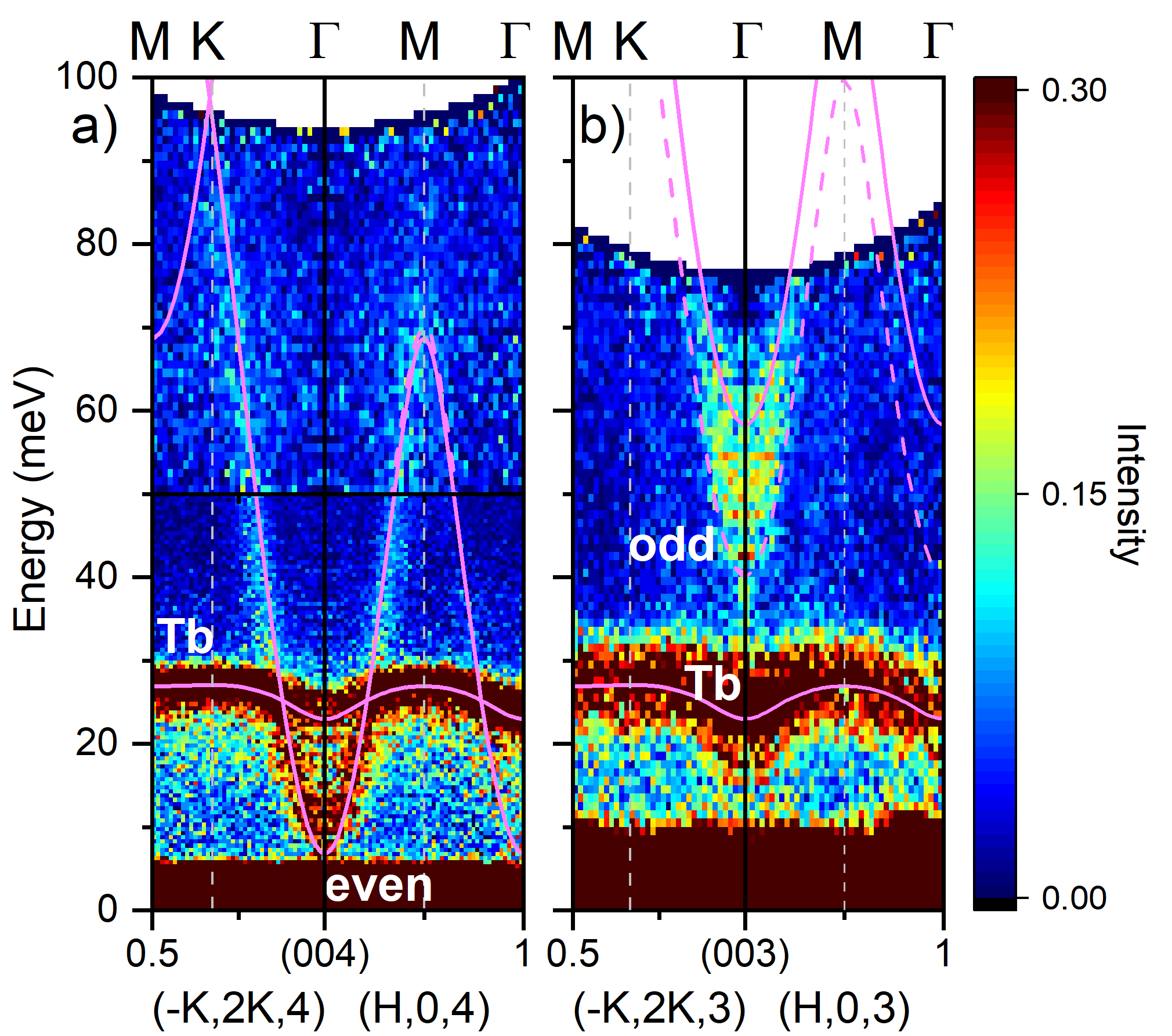}
\caption{\footnotesize (a) Slices of the data highlighting the dispersion of the even mode along the ($H$,0,0) and ($-K$,$2K$,0) directions in the (0,0,4) zone with $E_i=$ 160 meV (lower panel) and $E_i=$ 250 meV (upper panel).  (b) Slices of the data highlighting the dispersion of the odd mode along the ($H$,0,0) and ($-K$,$2K$,0) directions in the (0,0,3) zone with $E_i=$ 160 meV.  For (a) and (b), the data are averaged over $\Delta L=\pm 0.5$ and either $\Delta H=\pm 0.1$ or $\Delta K=\pm 0.058$.  In all panels, pink lines correspond to model dispersions with $L=0$ (solid lines) and $L=0.5$ (dashed lines).}
\label{fig_even_odd_intra}
\end{figure} 

\begin{figure}
\includegraphics[width=1.0\linewidth]{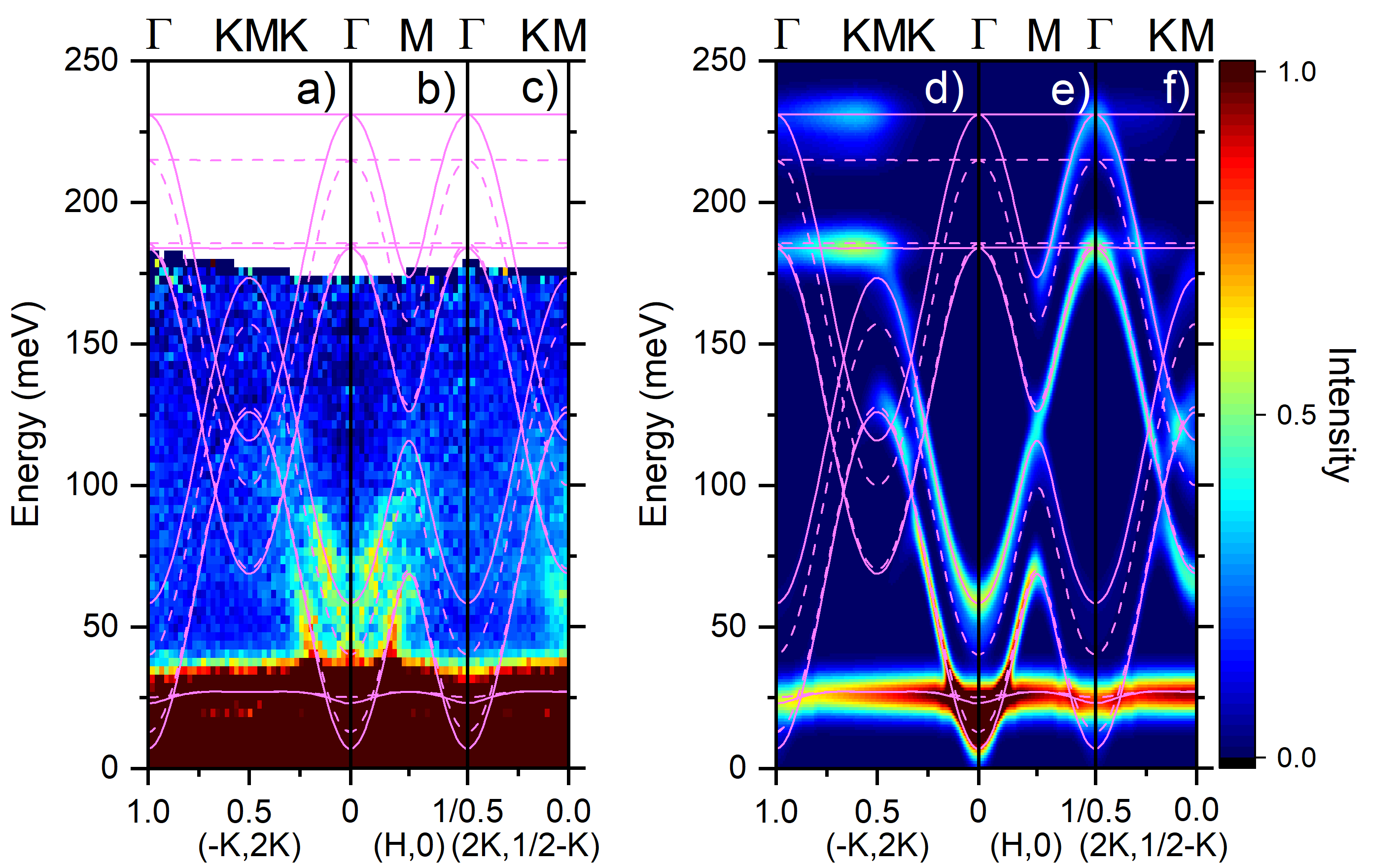}
\caption{\footnotesize Slices of the $E_i=250$ meV data after averaging over $\Delta L=\pm 7$ showing the dispersion along the (a) ($-K$,$2K$,0), (b) ($H$,0) and (c) ($2K$,$1/2-K$) directions. For all panels, the data are additionally averaged over either $\Delta H=\pm 0.1$ or $\Delta K=\pm 0.058$. (d)-(f) Model calculations of the neutron intensities with the same reciprocal space averaging of the data as in (a)-(c) and convolved with a Gaussian energy FWHM of 12 meV.  In all panels, pink lines correspond to model dispersions with $L=0$ (solid lines) and $L=0.5$ (dashed lines). }
\label{fig_high_energy}
\end{figure}

The intralayer dispersions are steeper than the interlayer modes and can extend well beyond 100 meV.
The odd and even modes can be isolated in the INS data based on their structure factors which are maximized in Brillouin zones with $L=odd$ and $L=even$, respectively.
Slices from the $E_i=$75 and 160 meV data corresponding to even modes with $L=4$ and odd modes with $L=3$ are shown in Figs.~\ref{fig_even_odd_intra}(a) and \ref{fig_even_odd_intra}(b), respectively.
To gain better statistics, the data are averaged over $\Delta L=\pm 0.5$ rlu which broadens features by effectively averaging over the interlayer bandwidth.
For $L=4$, the even mode has a M-point energy of $\approx$ 70 meV.
For $L=3$, the odd mode is more strongly broadened by interlayer interactions than the even mode, but we clearly observe the even-odd mode splitting of $\approx$ 55 meV.

We obtained the intralayer exchange parameters defined in $\mathcal{H}_{\rm intra}$ by fitting various cuts of the lowest odd and even branches similar to those shown in Figs.~\ref{fig_spin_gap} and \ref{fig_even_odd_intra}.
During the fit, all parameters of $\mathcal{H}_{\rm inter}$ and $\mathcal{H}_{\rm aniso}$ were fixed to the values in Table \ref{Heisenberg}.
Ultimately, we achieved satisfactory agreement with the data with only one parameter corresponding to the nearest-neighbor Mn-Mn intralayer FM interaction with $J=-28.8$(2) meV.
The main reason for this simple result is that the dispersive features quickly deteriorate at higher energies by becoming very broad and weak.

Figure \ref{fig_high_energy}(a)-(c) shows intralayer dispersion data after summing over a large range of $\Delta L=\pm 7$ rlu.
This improves statistics and allows higher-energy features to be observed, but it mixes odd and even modes and averages over the interlayer dispersions.
Excitations are observed up to $\sim$125 meV which includes evidence for the top of the odd branch near the M-point at $\sim$ 115 meV [Fig.~\ref{fig_high_energy}(b)] and the bottom of the fourth branch (even) at the M-point near 70 meV [Fig.~\ref{fig_high_energy}(c)].
These data are compared to model calculations in Figs.~\ref{fig_high_energy}(d)-(f) that average over the same reciprocal space ranges.
From the model, the K-point Dirac crossing of the even mode is predicted to occur near 90 meV.
However, we are not able to resolve any K-point gapping in the INS data.

\section{First-principles calculations of the intrinsic magnetic properties}

\begin{figure}[tbh]
	\begin{tabular}{c}
\includegraphics[width=1.0\linewidth,clip]{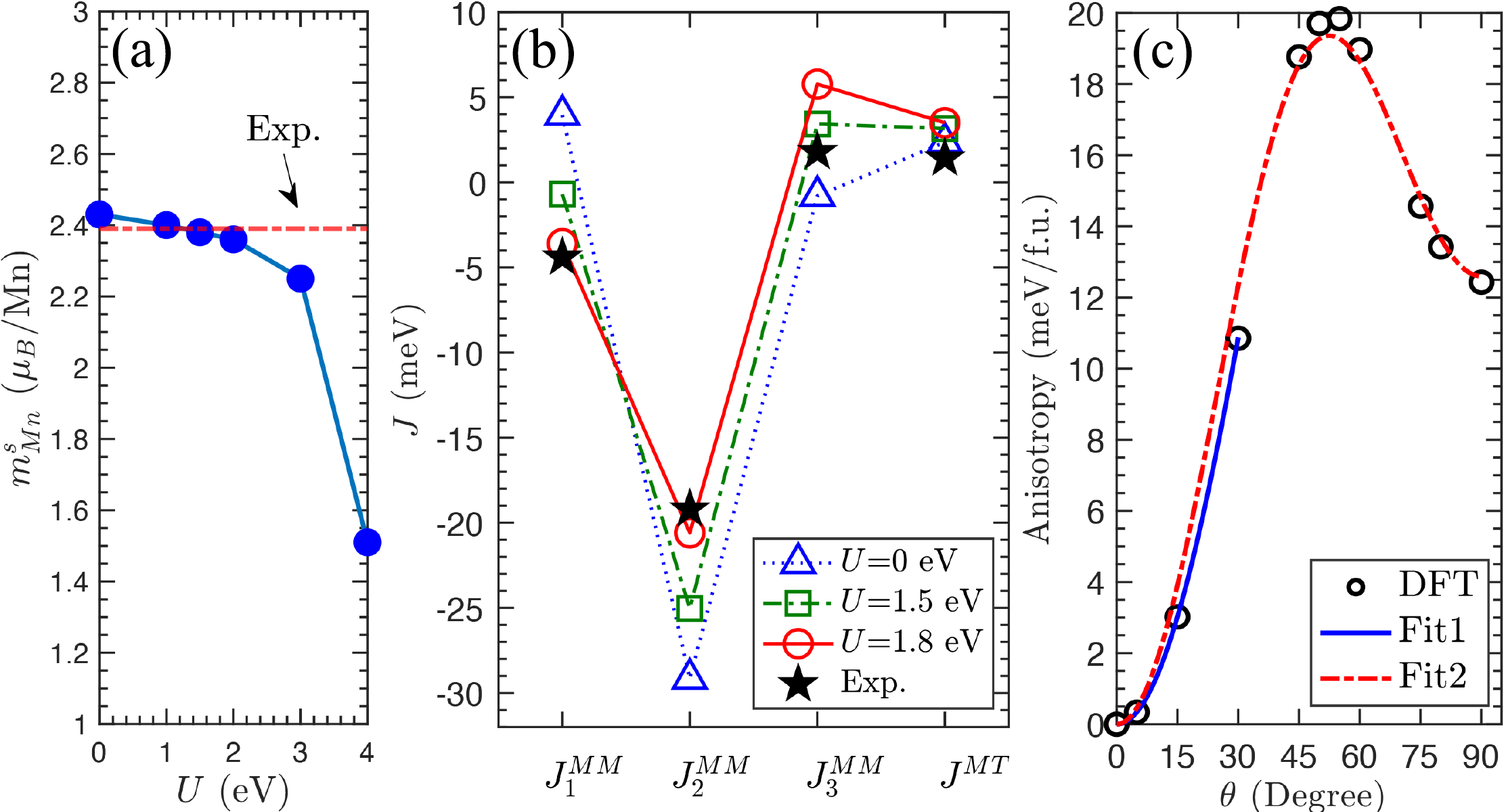}%
	\end{tabular}%
        \caption{ Intrinsic magnetic properties calculated in Tb166 and compared to the experimental values.
          (a) On-site Mn spin magnetic moment $m_{\rm Mn}^s$ and (b) interlayer exchange parameters as functions of Hubbard $U$ applied on Mn $3d$-states.
          Hubbard $U$ on Mn-$3d$ is included using the around-the-mean-field double-counting scheme in DFT+$U$.          
(c) Variation of energy as a function of spin-quantization axis rotation.
          $\theta=0\degree$ corresponds to the out-of-plane spin orientation parallel to the $c$-axis.
        Fit1 and Fit2 correspond to fittings to the expressions of $E(\theta)=K_1 \sin^2\theta$ near $\theta=0$ and $E(\theta)=K_1 \sin^2\theta + K_2 \sin^4\theta$ over the full $\theta$ range, respectively.
        }
	\label{DFT_fig}
\end{figure}

DFT calculations were carried out to investigate the intrinsic magnetic properties in Tb166, which includes magnetization, the interlayer exchange couplings, and magnetocrystalline anisotropy (MA).  
The  strongly correlated Tb-4$f$ states were treated in both the DFT+$U$ method and the so-called open-core approach.
We also explored the effects on the exchange couplings of additional electron repulsion for Mn-$3d$ orbitals in DFT+$U$.
Details of these calculations can be found in the SM~\cite{SI}.
Results are shown in \rfig{DFT_fig}.

We first investigate the spin and orbital magnetic moments in Tb166.
Tb-$4f$ are treated within DFT+$U$ using the fully-localized-limit (FLL) double-counting scheme, and spin-orbit-coupling (SOC) is included using the second variation method.
The calculated spin and orbital magnetic moments of Tb are $m_s^{\rm Tb}=6.26~\mub$ and $m_l^{\rm Tb}=2.96~\mub$, respectively, consistent with Hund's rules.
The calculated total magnetic moments of Tb and Mn, $m^{\rm Tb}=$ 9.23~$\mub$ and $m^{\rm Mn}=$2.42~$\mub$, respectively, agree with the low-temperature experimental results of $m^{\rm Tb}=9.0~\mub$ and $m^{\rm Mn}=2.17~\mub$~\cite{Clatterbuck99}.

The four interlayer isotropic exchange couplings discussed above are calculated by mapping the total energies of five collinear spin configurations (see SM \cite{SI}) into $\mathcal{H}_{\rm inter}$ defined in Eqn.~(\ref{Hinter}).
Mn and Tb spin derived from the spin magnetic moment, $s_{\rm Mn} = m_s^{\rm Mn}/2$ and $S_{\rm Tb} = m_s^{\rm Tb}/2$, are used in the mapping procedure.
The overall ferrimagnetic structure is stabilized by $\jmmb$ and $\jmt$.
In all our calculations, we found that the Mn-Tb coupling $\jmt$ is AFM and is a strong contributor to the overall magnetic energy when considering the high Tb spin and multiplicity of 12 neighboring Mn atoms.
The dominant interlayer Mn-Mn coupling, the FM $\jmmb$, is also confirmed in DFT, although its amplitude is overestimated by $\sim50\%$.
On the other hand, we found AFM $\jmma$ and FM $\jmmc$.
All three calculated $\jmmk$ have the same sign as the values calculated for Y166~\cite{Dally21}, and their amplitudes are also comparable~\cite{Ghimire20}.
However, for the weaker couplings $\jmma$ and $\jmmc$, the signs of calculated values disagree with those deduced from INS.

To resolve this discrepancy, we consider the electron correlation effects of Mn-$3d$ orbitals on exchange couplings in DFT+$U$.
We note that various $U$ values have been applied on Mn-$3d$ orbitals in the previous studies of R166.
For example, Tb166 bandstructure was calculated in plain DFT ($U=0$) while $U=\SI{4}{\eV}$ was used in DFT+DMFT to explain the band structures of Y166 measured by ARPES~\cite{Li21}.
Especially, the $U$ dependence of interlayer Mn-Mn couplings in Y166 has already been investigated with $U=\SIrange{0}{3.5}{\eV}$ using the FLL double-counting scheme in DFT+$U$.
However, as shown in Ref.~[\onlinecite{Ghimire20}], the FLL scheme quickly overestimates the Mn magnetic moment with finite $U$.
Thus, instead, here we use the around-the-mean-field double-counting scheme~\cite{Czyzyk_1994}, which is usually believed to be more suitable for less-strongly-correlated metallic systems.
Unlike the FLL scheme, we found that the $m^{\rm Mn}_s$ remains close to experimental value with $U$ values of 0--2~eV, as shown in \rfig{DFT_fig}(a).
Compared to magnetization, the variation of the exchange parameters is much more pronounced, although the experimental state has the lowest energy for $U=\SIrange{0}{2}{\eV}$.
\rFig{DFT_fig}(b) shows $J_i^\text{MM}$ ($i=1,2,3$) and $\jmt$ calculated using various $U$ values, compared to experiment.
Remarkably, both $\jmma$ and $\jmmc$ can change their signs with increasing $U$.
With $U=\SIrange{1.5}{1.8}{\eV}$, the signs of all interlayer $J$ values become consistent with those deduced from INS.
Thus, while DFT gives a reasonable description of the dominant magnetic interactions in Tb166, including Mn-$3d$ electron correlations can further improve the description of $\jmma$ and $\jmmc$.

Electron correlations can have profound effects on magnetic interactions and spin excitations~\cite{ke2021ncm}, especially in more localized systems.
The most recent Mn-based examples include the extensively studied layered topological materials, MnBi$_2$Te$_4$~\cite{Li_2020} and MnSb$_2$Te$_4$~\cite{Lai_2021,Riberolles_2021}, where a sizable $U=\SIrange{4}{5}{\eV}$ on Mn-$d$ orbitals was needed to correctly describe the magnetic interactions in DFT+$U$ while the plain DFT fails to predict the correct magnetic ground state.
Although the widely-used DFT+$U$ method provides the simplistic Hubbard correction beyond DFT, the choice of the correlated orbitals and the associated value of the Hubbard $U$ parameter is not well-defined for metallic systems like Tb166.
Moreover, the non-local exchange-correlation potentials can also be important, and a simple $U$ parameter may not be sufficient~\cite{lee2020prb} to best describe the electronic structures.
Future experimental and theoretical works may be helpful to further clarify the electron correlation role in Tb166 and determine the best $U$ parameter.

The MA energy (MAE) is also investigated by calculating the total energies of the ferrimagnetic state as a function of spin-quantization direction, which is shown in \rfig{DFT_fig}(c).
In agreement with the ground state structure of Tb166, the MAE displays strong uniaxial anisotropy with a minimum energy at $\theta=0$ (easy-axis) relative to the $c$-axis.
Moreover, the non-monotonic dependence of $E$ on $\theta$ is consistent with substantial higher-order MAE constants.
Over the full range of $\theta$, we fit MA energy (see Fit2 in \rfig{DFT_fig}(c)) to the expression
\begin{equation}
E(\theta)=K_1 \sin^2\theta + K_2 \sin^4\theta.
\end {equation}
The resulted large ratio of $K_2/K_1=-1.25$ is sufficient to drive the spin reorientation transition~\cite{Zajkov00}.
To better compare with the single-ion anisotropy deduced from low-temperature INS, we also fit MA energy (see Fit1 in \rfig{DFT_fig}(c)) with $E(\theta)=K_1 \sin^2\theta$ near $\theta=0$, which corresponds to the ground state anisotropy.
This provides $K_1\approx 43$~meV/f.u. and can be compared to our experimental value [see Eqn.(\ref{Haniso}) and Table~\ref{Heisenberg}] according to  $K_{\rm tot}=-(K^{\rm T} S^2+6K^{\rm M}s^2)=23.1$~meV/f.u.
Thus, DFT overestimates the MAE by $\sim85\%$, which is a reasonable agreement considering that an accurate ab initio description of MA is generally challenging, especially in complex $4f$ intermetallics.
The Tb-$4f$ contributions dominate the easy-axis MA in Tb166 as the Mn sublattice contribution is one-order of magnitude smaller and easy-plane.

\section{Discussion}
The INS data for Tb166 provide a minimal set of exchange and anisotropy parameters that are largely consistent with our DFT results and indirect estimations of these energy scales from magnetization and neutron diffraction data (see e.g. Refs.~\cite{Venturini96,Zajkov00,Rosenfeld08}). The key conclusions are: (1) large intralayer FM interactions between Mn ions, (2) interlayer interactions that are dominated by FM coupling between Mn layers spaced by Sn layers ($J^{\rm MM}_2$) and AFM coupling between Mn and Tb layers, (3) the presence of competing, weaker AFM and FM Mn-Mn interlayer couplings, and (4) a net uniaxial magnetic anisotropy.

With respect to (2) and (3), we consider the overall stability of the ferrimagnetic structure of Tb166 by examining the classical magnetic energies of collinear layer stackings given by
\begin{multline}
E = 6J^{\rm MT}[(\bold{s}_1+\bold{s}_2)\cdot\bold{S}_a+(\bold{s}_3+\bold{s}_4)\cdot\bold{S}_b]+3J^{\rm MM}_1(\bold{s}_1\cdot\bold{s}_2+\bold{s}_3\cdot\bold{s}_4) 
\\ + 3J^{\rm MM}_2(\bold{s}_1\cdot\bold{s}_4+\bold{s}_2\cdot\bold{s}_3) + 6J^{\rm MM}_3(\bold{s}_1\cdot\bold{s}_3+\bold{s}_2\cdot\bold{s}_4).
\end{multline}
Here, the numbers label successive Mn layers and letters label Tb layers for a six-layer stack. The ground state ferrimagnetic structure has an energy of 
\begin{equation}
E_{\rm ferri} = -24sSJ^{\rm MT}-6s^2(J^{\rm MM}_1+J^{\rm MM}_2-2J^{\rm MM}_3).
\end{equation}
The next higher-energy state corresponds to AFM up-down-down-up (UDDU) Mn layer stacking.  For uniaxial anisotropy, the classical UDDU state will decouple the Mn and the Tb layers and
\begin{equation}
E_{\rm UDDU} = -6s^2(-J^{\rm MM}_1+J^{\rm MM}_2+2J^{\rm MM}_3).
\end{equation}
The parameters in Table \ref{Heisenberg} provide $E_{\rm ferri}=-$ 220 meV and $E_{UDDU}=-$110 meV, indicating that the high stability of the ferrimagnetic ground state arises from $J^{\rm MT}$.  In the absence of $J^{\rm MT}$ (as for Y166), the collinear ferromagnetic, ferrimagnetic and UDDU states are nearly degenerate since $J^{\rm MM}_1 \approx -2J^{\rm MM}_3$.  This suggests that similar competition between these interlayer interactions drives complex helical ordering observed in Y166.

Based on these comparisons, it is interesting to consider the transferability of exchange interactions in Tb166 with other $R$166 compounds.  INS investigations of Y166 in Ref.~\cite{Zhang20} report a NN intralayer exchange that is nearly identical to Tb166.  While the interlayer interactions in Y166 are not studied in detail in Ref.~\cite{Zhang20}, the bilayer splitting energy is reported as $|J^{\rm MM}_1+J^{\rm MM}_2|\approx$24 meV, which is the same as Tb166.  This suggests that $J^{\rm MM}_1$ and $J^{\rm MM}_2$ interactions are both FM and have similar strengths in Y166 and Tb166. One caveat is that additional intralayer and interlayer interactions are also fit in Ref.~\cite{Zhang20}. Interestingly, our DFT calculations support an AFM $J^{\rm MM}_1$ and FM $J^{\rm MM}_3$, and vice versa, with the result depending on the choice of the correlation parameter $U$. Overall, these comparisons give some confidence that the Mn-Mn magnetic interactions in $R$166 compounds share a remarkable similarity: the $J^{\rm MM}_2$ is FM and dominates the interlayer Mn-Mn coupling, while $J^{\rm MM}_1$ and $J^{\rm MM}_3$ are much weaker and competing. The variation of $R$ ion and slight changes in structure will likely affect the overall balance of $J^{\rm MM}_1$ and $J^{\rm MM}_3$.

There is little data reporting the magnitude of the Mn-$R$ coupling in other $R$166 compounds.
For Gd166, the energy scale for the Gd mode is reported to be $\sim$24 meV from powder INS data \cite{Tils98} which is very similar to the Tb mode energy observed here.
However, given the absence of Gd single-ion anisotropy, simulations (see SM \cite{SI}) show that this energy corresponds to the top of the Gd mode at $\approx \omega_F+2JS^{\rm Gd}=12sJ^{\rm MnGd}$, allowing an estimate of $J^{\rm MnGd} \approx 2$ meV.
The energy of the Tb mode is lifted appreciably by anisotropy, $2SK^{\rm T}=10$ meV.
Thus, our reported $J^{\rm MT}$ is about 30\% smaller than $J^{\rm MnGd}$, a result that is roughly consistent with a decrease of $4f$-$5d$ overlap due to lanthanide contraction \cite{Lee_22}.
Extrapolating to Ho166 and Er166 should result in weaker ferrimagnetism.
For Er166, this weakening results in the observed decoupling of the Mn and Er sublattice magnetic ordering at high temperatures \cite{Venturini91}.

The magnetic anisotropies of Tb166 determined from INS may present some inconsistencies with our understanding of $R$166 compounds.  At low temperatures, Tb166 is dominated by the large uniaxial anisotropy of the Tb ion, a result that is consistent with our INS data and DFT results.  However, the INS data cannot be modeled with an easy-plane Mn anisotropy parameter since the spin gap becomes too small.  Instead, we obtain the best fitting results by assuming that Mn also has uniaxial single-ion anisotropy.  This is inconsistent with INS data from Y166 that finds a rather large value of 5 meV for the Mn easy-plane single-ion anisotropy parameter, although the spin gap itself is not reported \cite{Zhang20}.  It is very possible that both Mn-Mn and Mn-Tb exchange anisotropy contributes to the spin gap as well.   First-principles calculations find significant exchange anisotropy of the intralayer coupling in Y166 \cite{Ghimire20}.  In Tb166, the Tb magnetic anisotropy is temperature dependent, and our MAE calculations in the ground state are consistent with the expected conditions for the spin reorientation transition that occurs at 350~K. It will be interesting to study the spin excitations in this temperature regime to learn more about the unusual magnetic anisotropy of $R$166 compounds.

Finally, we would like to discuss briefly the role that magnetic instabilities and fluctuations play in the band topology of Tb166. The magnetic stacking of FM Mn and Tb layers in Tb166 is very stable to competing interlayer interactions due to the large Tb-Mn coupling. Thus, the only  avenue available for tuning of topological band states in Tb166 is by controlling the magnetic anisotropy and, consequently, the spin reorientation transition. This will affect the size of the Chern gap, which is maximized for the uniaxial moment configuration. On the approach to the spin reorientation at elevated temperatures, we might ask whether magnetic fluctuations play any role in quantum transport. Recent muon spectroscopy results report a correlation between quantum transport in Tb166 and the suppression of slow ($\sim$MHz) magnetic fluctuations that appear below 120 K \cite{Mielke21}. The origin of these slow magnetic fluctuations is a mystery, but our INS data indicate that they do not arise from collective spin wave modes which are gapped out on a THz scale.

\section{Summary}
INS data for Tb166 provide a minimal set of exchange and anisotropy parameters that are largely consistent with indirect estimations of these energy scales provided by magnetization data and neutron diffraction, as well as by our DFT calculations.  The key conclusions are: (1) large intralayer FM interactions between Mn ions, (2) interlayer interactions that are dominated by FM coupling between Mn layers spaced by Sn layers ($J^{\rm MM}_2$) and AFM coupling between Mn and Tb layers, (3) the presence of weaker FM and AFM Mn-Mn interlayer couplings, and (4) an overall uniaxial magnetic anisotropy.  These results suggest that the magnetism of $R$166 compounds, with a variety of magnetic ground states and high-temperature or high-field instabilities, may be understood with transferable set of magnetic interactions. A complete understanding of these interactions and their evolution through the R166 family could allow for the prediction of additional topological responses accessible via tuning of the magnetism using external applied fields or rare-earth engineering protocols.

\section{Acknowledgments} RJM, LK, YL, BGU, BL and SXMR's work at the Ames Laboratory is supported by the U.S. Department of Energy (USDOE), Office of Basic Energy Sciences, Division of Materials Sciences and Engineering. TJS and PC are supported by the Center for the Advancement of Topological Semimetals (CATS), an Energy Frontier Research Center funded by the USDOE Office of Science, Office of Basic Energy Sciences, through the Ames Laboratory.  Ames Laboratory is operated for the USDOE by Iowa State University under Contract No. DE-AC02-07CH11358. TJS is also partially funded by the Gordon and Betty Moore Foundation (Grant No. GBMF4411).  A portion of this research used resources at the Spallation Neutron Source, which is a USDOE Office of Science User Facility operated by the Oak Ridge National Laboratory. L.K. is supported by the U.S. DOE, Office of Science, Office of Basic Energy Sciences, Materials Sciences and Engineering Division, and Early Career Research Program. A portion of this research used resources of the National Energy Research Scientific Computing Center (NERSC), a U.S. DOE Office of Science User Facility operated under Contract No. DE-AC02-05CH11231

\end{document}